\documentclass[12pt]{article}
\usepackage{natbib}
\usepackage{graphicx}
\usepackage{amsmath}
\usepackage{amssymb}
\usepackage{authblk}
\usepackage{Sweave}
\usepackage{hyperref}

\newcommand{\proglang}{\texttt}
\newcommand{\pkg}{\texttt}
\newcommand{\code}{\texttt}

\begin{document}

\title{Robust Estimation of the Generalized Loggamma Model. The \proglang{R} Package \pkg{robustloggamma}}

\author[1]{Claudio Agostinelli\thanks{Corresponding author. claudio.agostinelli@unitn.it}} 
\author[2]{A. Marazzi}
\author[3]{V.J. Yohai}
\author[2]{A. Randriamiharisoa}

\affil[1]{Department of Mathematics, University of Trento, Trento, Italy}
\affil[2]{Institute of social and preventive medicine,
 Lausanne University Hospital, Switzerland}
\affil[3]{Departamento de Matematicas, Facultad de Ciencias Exactas y Naturales, University of Buenos Aires, Argentina}

\date{1 October 2015}

\maketitle

\begin{abstract}
\pkg{robustloggamma} is an \proglang{R} package for robust estimation and inference in the generalized loggamma model. We briefly introduce the model, the estimation procedures and the computational algorithms. Then, we illustrate the use of the package with the help of a real data set.
  
\end{abstract}

\newpage

\section{Introduction}
\label{sec:intro}

Generalized loggamma distribution is a flexible three parameter family introduced by \citet{stacy1962} and further studied by \citet{prentice1974} and \citet{lawless1980}. This family is used to model highly skewed positive data on a logarithmic scale and it includes several asymmetric families such as logexponential, logWeibull, and loggamma distributions and the normal distribution too. In the parametrization given by \citet{prentice1974} the three parameters are location $\mu$, scale $\sigma$, and shape $\lambda$. We denote the family by $LG(\mu, \sigma, \lambda)$, $\mu \in \mathbb{R}$, $\sigma >0$, $\lambda \in \mathbb{R}$. If $y$ is a random variable with distribution $LG(\mu, \sigma, \lambda)$ then $y$ is obtained by location and scale transformation
\begin{equation*}
y=\mu +\sigma u
\end{equation*}
of the random variable $u$ with density
\begin{equation*}
f_{\lambda}(u) = \left\{
\begin{array}{lll}
\frac{\left\vert \lambda \right\vert}{\Gamma \left( \lambda^{-2} \right)} (\lambda^{-2})^{\lambda^{-2}} \exp \left( \lambda^{-2} \left( \lambda u - e^{\lambda u} \right) \right) & \text{if} & \lambda \neq 0, \\ 
\frac{1}{\sqrt{2\pi}} \exp(-\frac{u^2}{2}) & \text{if} & \lambda = 0 \ ,
\end{array}
\right.
\end{equation*}
where $\Gamma$ denotes the Gamma function. Hence, the density of $y$ is $f_{\boldsymbol{\theta}}(y) = f_{\lambda} \left( (y - \mu)/\sigma \right)/\sigma$ where $\boldsymbol{\theta} = (\mu, \sigma, \lambda)$. Normal model ($\lambda=0$), logWeibull model ($\lambda=1$), logexponential model ($\lambda=1$ and $\sigma=1$), and loggamma model ($\sigma=\lambda$) are special cases.  The \textit{generalized gamma} family is obtained by back transforming on the original scale, i.e., $t=\exp(y)$; in this situation the expectation is $\eta = E(t) = \delta \Gamma(\alpha + 1/\gamma) / \Gamma(\alpha)$ where $\alpha = \lambda^{-2}$, $\gamma = \lambda /\sigma$, $\delta =\exp(\mu +2\log(\lambda) \sigma /\lambda)$ is an important parameter.

\begin{figure}
  \centerline{
     \includegraphics[width=0.7\textwidth]{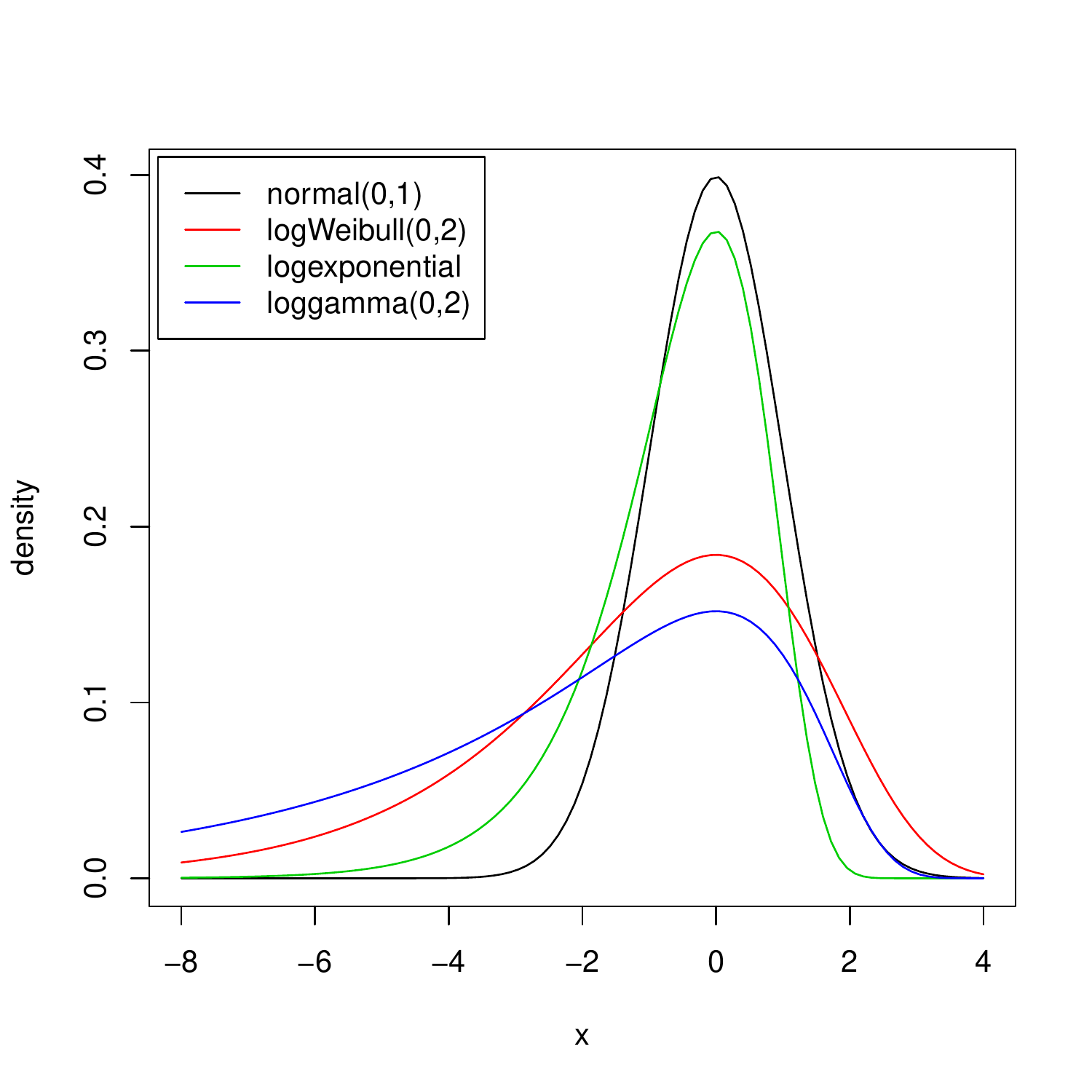}
  }
  \caption{Density for some relevant members of the generalized loggamma family.}
  \label{fig:loggamma}
\end{figure}

The \pkg{robustloggamma} package provides density, distribution function, quantile function and random generation for the loggamma distribution using the common syntax \code{[dpqr]loggamma}. In Figure~\ref{fig:loggamma} we draw the density of some relevant distributions using the following code.

\begin{Schunk}
\begin{Sinput}
R> require("robustloggamma")
R> plot(function(x) dloggamma(x, mu=0, sigma=1, lambda=0), 
+      from=-8, to=4, ylab="density")
R> plot(function(x) dloggamma(x, mu=0, sigma=2, lambda=1), 
+      from=-8, to=4, add=TRUE, col=2)
R> plot(function(x) dloggamma(x, mu=0, sigma=1, lambda=1), 
+      from=-8, to=4, add=TRUE, col=3)
R> plot(function(x) dloggamma(x, mu=0, sigma=2, lambda=2), 
+      from=-8, to=4, add=TRUE, col=4)
R> legend("topleft", legend=c("normal(0,1)", "logWeibull(0,2)", 
+   "logexponential", "loggamma(0,2)"), col=1:4, lty=1, inset=0.01)
\end{Sinput}
\end{Schunk}

\section{Robust estimation and inference}
We condider the three parameter family $LG(\mu, \sigma, \lambda)$ with $\boldsymbol{\theta}=(\mu, \sigma , \lambda)$, and distribution function $F_{\boldsymbol{\theta}}(y)=F^{\ast}((y-\mu)/\sigma, \lambda)$. For $0<u<1$ let $Q(u, \boldsymbol{\theta})$ be the $u$-quantile of $F_{\boldsymbol{\theta}}(y)$. Then, $Q(u, \boldsymbol{\theta}) = \sigma Q^\ast(u, \lambda) + \mu$, where $Q^\ast(u, \lambda) = Q(u, (0,1,\lambda))$. Let $y_{(1)}, \cdots, y_{(n)}$ be the order statistics of a sample of size $n$ from $LG(\mu_0, \sigma_0, \lambda_0)$, and $\boldsymbol{\theta}_0=(\mu_0, \sigma_0, \lambda_0)$ is the unknow vector of parameters to be estimated. Since, $y_{(j)}$ is the quantile $u_{n,j}=(j-0.5)/n$ of the empirical distribution, it should be close to the corresponding theoretical quantile $\sigma_0 Q^{\ast}(u_{n,j}, \lambda_0) + \mu_0$. Hence, the residuals:
\begin{equation*}
r_{n,j}(\boldsymbol{\theta}) = y_{(j)} - \mu - \sigma Q^\ast(u_{n,j}, \lambda)
\end{equation*}
are a function of $\boldsymbol{\theta}$ and they should be as small as possible. To summarize their size a \textit{scale} $s$ is often used. Given a sample $\mathbf{u} = (u_1, \cdots, u_n)$, $s$ is a function of $\mathbf{u}$ with two basic properties: (i) $s(\mathbf{u}) \geq 0$; (ii) for any scalar $\gamma$, $s(\gamma \mathbf{u}) = |\gamma| s(\mathbf{u})$. The most common scale is $\left(\sum_{j=1}^n u_j^2/n \right)^{1/2}$ and it is clearly non robust. To the aim of gaining robustness we use a $\tau$ scale $\tau(r_{n,1} (\boldsymbol{\theta}), \cdots, r_{n,n}(\boldsymbol{\theta}))$ introduced by \citet{yohaizamar1988}. A short review of $\tau$ scales can be found in Appendix~\ref{app:tauscaleregression}.

Then, the Q$\tau$ \textit{estimator} is defined by 
\begin{equation*}
\tilde{\boldsymbol{\theta}} = \arg\min_{\boldsymbol{\theta}} \tau\left( r_{n,1}(\boldsymbol{\theta}), \cdots, r_{n,n}(\boldsymbol{\theta}) \right) \ .
\end{equation*}
We note that, fixing $\lambda$, the value of $\mu$ and $\sigma$ minimizing the $\tau$ scale are obtained by a simple regression $\tau$ estimate for the responses $y_{(j)}$ and the regressors $Q^{\ast}(u_{n,j}, \lambda)$. We also note \citep{serfling1980} that $n^{1/2} r_{n,j}(\boldsymbol{\theta}_0)$ is approximately distributed according to $N(0,v^{2}(\boldsymbol{\theta}_0, u_{n,j}))$, where
\begin{equation*}
v^2(\boldsymbol{\theta}_0, u) = \frac{\sigma_0^2 u (1-u)}{f_{\lambda_0}^2(Q^{\ast}(u, \lambda_0))} \ .
\end{equation*}
Then, the variances of the regression errors can be estimated by
\begin{equation*}
\tilde{\sigma}_j^2 = v^2(\tilde{\boldsymbol{\theta}}_n, u_{n,j})
\end{equation*}
and the basic estimator can be improved by means of a weighted procedure.
More precisely, one defines the \textit{weighted} Q$\tau$ \textit{estimator} (WQ$\tau$), with the set of weights $1/\tilde{\sigma}_1, \cdots, \tilde{\sigma}_n$, by
\begin{equation*}
\tilde{\boldsymbol{\theta}}^w = \arg\min_{\boldsymbol{\theta}} \tau \left( \frac{r_{n,1}(\boldsymbol{\theta})}{\tilde{\sigma}_1}, \cdots, \frac{r_{n,n}(\boldsymbol{\theta})}{\tilde{\sigma}_n} \right) \ .
\end{equation*}

Monte Carlo simulations \citep{tech2013} show that both the Q$\tau$ and the WQ$\tau$ estimators perform well in the case the model is correct and also when the sample contains outliers. These empirical findings are corroborated by a theoretical results showing that Q$\tau$ and WQ$\tau$ have a $50\%$ break down point (BDP) (according to a special definition of BDP - the finite sample distribution break down point - which is particularly designed to asses the degree of global stability of a distribution estimate).

\subsection{Weighted likelihood estimators}

Unfortunately, Q$\tau$ and WQ$\tau$ are not asymptotically normal and therefore inconvenient for inference. Their rates of convergence is however of order $n^{1/2}$ and this makes them a good starting point for a one-step weighted likelihood (WL) procedure which is asymptotically normal and fully efficient at the model. The package \pkg{robustloggamma} implements two WL estimators: the fully iterated weighted likelihood and the one step weighted likelihood. Monte Carlo simulations \citep{tech2013} show that both these estimators maintain the robust properties (BDP) of Q$\tau$ and WQ$\tau$. 
 
In general, a \textit{weighted likelihood estimator} (WLE) as defined in \citet{markatoubasulindsay1998} is a solution of the following estimating equations
\begin{equation*}
\frac{1}{n} \sum_{j=1}^{n} w(y_{j},\boldsymbol{\theta}) \mathbf{z}(y_{j},\boldsymbol{\theta}) = \mathbf{0} \ ,
\end{equation*}
where $\mathbf{z}(y,\boldsymbol{\theta})$ is the usual score function vector and $w(y, \boldsymbol{\theta})$ is a weight function defined by
\begin{equation*}
w(y, \boldsymbol{\theta}) = \min \left( 1, \frac{\left[ A(\delta(y, \boldsymbol{\theta})) + 1 \right]^{+}}{\delta(y, \boldsymbol{\theta}) + 1} \right) \ ,
\end{equation*}
where $\delta(y, \boldsymbol{\theta})$ is the \textit{Pearson residual} \citep{lindsay1994}, measuring the agreement between the distribution of the data and the assumed model. It is defined as $\delta(y, \boldsymbol{\theta}) = \left[ f^{\ast}(y) - f_{\boldsymbol{\theta}}^{\ast}(y) \right] / f_{\boldsymbol{\theta}}^{\ast}(y)$, where $f^{\ast}(y) = \int k(y,t,h) \ dF_{n}(t)$ is a kernel density estimate of $f_{\boldsymbol{\theta}}$ (with bandwidth $h$), $f_{\boldsymbol{\theta}}^{\ast}(y)=\int k(y,t,h) f_{\boldsymbol{\theta}}(t) \ dt$ is the corresponding smoothed model density, $F_{n}$ is the empirical cumulative distribution function, and $[x]^{+}=\max (0,x)$. 

The function $A(\cdot)$ is called \textit{residual adjustment function} (RAF). When $A(\delta)=\delta$ the weights $w(y_{j},\boldsymbol{\theta})=1$ and the WLE equations coincides with classical MLE equations. Generally, the weight function $w$ uses RAF that correspond to minimum disparity problems \citep{lindsay1994}, see Appendix~\ref{app:RAF} for some examples.

The \textit{fully iterated weighted likelihood estimator} (FIWL) is the solution of the weighted equations, while the \textit{one-step weighted likelihood estimator} (1SWL) is defined by
\begin{equation*}
\hat{\boldsymbol{\theta}} = \tilde{\boldsymbol{\theta}} - \mathbf{J}^{-1} \sum_{j=1}^{n} w(y_{j}, \tilde{\boldsymbol{\theta}}) \mathbf{z}(y_{j}, \tilde{\boldsymbol{\theta}}) \ ,
\end{equation*}
where $\mathbf{J} = \int w(y, \tilde{\boldsymbol{\theta}}) \nabla \mathbf{z}(y, \tilde{\boldsymbol{\theta}}) \ dF_{\tilde{\boldsymbol{\theta}}}(y)$ and $\nabla$ denotes differentiation with respect to $\boldsymbol{\theta}$. This definition is similar to a Fisher scoring step, where an extra term obtained by differentiating the weight with respect to $\boldsymbol{\theta}$ is dropped since, when evaluated at the model, is equal to zero. Further information on minimum distance methods and weighted likelihood procedures are available in \citet{basushioyapark2011}.

\section{Algorithms and implementation}
In the following sections, we describe the computation of the estimators implemented in the main function \code{loggammarob}. We first recall its arguments, a reference chart is reported in the Appendix~\ref{app:help}. The only required argument is \code{x} which contains the data set in a numeric vector. The argument \code{method} allows to choose among the available robust procedures. The default method is \code{"oneWL"}, a one step weighted likelihood estimator starting from WQ$\tau$. Other alternatives are \code{"QTau"} (Q$\tau$), \code{"WQTau"} (WQ$\tau$), \code{"WL"} (Fully iterated weighed likelihood) and \code{"ML"} (Maximum likelihood). When \code{method} is not  Q$\tau$ an optional numeric vector of length $3$ (location, scale, shape) could be supplied in the argument \code{start} to be used as starting value, otherwise, the default is WQ$\tau$ for the likelihood based methods and Q$\tau$ for WQ$\tau$. By default, weights in the WQ$\tau$ are specified as described in the previous section, if a different set of weights are needed the \code{weights} argument could be used. Fine tuning parameters are set by the function \code{loggammarob.control} and passed to the main function by the \code{control} parameter.

\subsection[Computation of Qtau and WQtau]{Computation of Q$\tau$ and WQ$\tau$}
To optimize the $\tau$ scale for a given value of $\lambda$, \pkg{robustloggamma} uses the resampling algorithm described in \citet{salibianbarrerawillemszamar2008}. Let $x_j = Q^\ast(u_{n,j},\lambda)$ and consider the following steps:
\begin{enumerate}
\item Take a random subsample of size $2$ made of the pairs $(x_{(j_1)}, y_{(j_1)})$ and $(x_{(j_2)}, y_{(j_2)})$.
\item Compute a preliminary estimate of $\mu$ and $\sigma$ of the form
\begin{equation*}
\sigma^{(0)} = \frac{y_{(j_1)} - y_{(j_2)}}{x_{(j_1)} - x_{(j_2)}} \qquad \qquad \mu^{(0)} = y_{(j_1)} - \sigma_i^{(0)} x_{(j_1)}
\end{equation*}
\item Compute the residuals $r_{j}^{(0)} = y_{(j)} - \sigma^{(0)} x_j - \mu^{(0)}$ for $j=1,...,n$. 
\item Compute least squares estimates $\mu^{(1)}$, $\sigma^{(1)}$ based on the $n/2$ pairs with the smallest absolute residuals $r_{j}^{(0)}$.
\item Compute the residuals $r_{j}^{(1)}$ for $j=1,...,n$ and the $\tau$ scale  $\tau( r_1^{(1)}, \cdots, r_n^{(1)} )$.
\end{enumerate}
Steps 1-5 are repeated a large number $N$ of times and the values $\mu^{(1)}$, $\sigma^{(1)}$ corresponding to the minimal $\tau$ scale are retained. These values are then used as starting values of an IRWLS algorithm, where the weights are defined, at each iteration, as $w_j = W \phi_{j1} + \phi_{j2}$ and
\begin{align*}
W   & = \frac{\sum_{j=1}^n 2 \rho_2(r_j/s) - \psi_2(r_j/s) r_j/s}{\sum_{j=1}^n \psi_1(r_j/s) r_j/s}, \\
\phi_{jk} & = \psi_k(r_j/s) / (r_j/s), \qquad k = 1,2 \ ,
\end{align*}
$\psi_k$ is the first derivative of the $\rho$ function $\rho_k$ (see Appendix~\ref{app:tauscaleregression}), and $s$ is a scale which is recursively updated as follows
\begin{equation*}
s = s \left( \frac{2}{n} \sum_{j=1}^n \rho_1(r_j/s) \right)^{1/2}
\end{equation*}
with initial value $s=\text{median}(|r_j|)/0.6745$.

This algorithm is used to compute $\mu(\lambda)$ and $\sigma(\lambda)$ for all values of $\lambda$ in a given grid $\lambda_1, \cdots, \lambda_k$. The final value of $\lambda$ is then obtained by minimizing the $\tau$ scale over the grid. 

The first part of the algorithm is implemented in \proglang{Fortran} while the IRWLS algorithm is implemented in \proglang{R}. The initial random procedure uses the \proglang{R} uniform pseudo random number generator and can be controlled by setting the seed in the usual way. The function \code{loggammarob.control} is used to set all the other parameters. \code{tuning.rho} and \code{tuning.psi} set the constants $c_1$ and $c_2$ of the $\rho$ functions $\rho_1$ and $\rho_2$.
\code{nResample} controls the number of subsamples $N$, \code{max.it}, and \code{refine.tol} provide the maximum number of iterations and the tolerance of the IRWLS algorithm. An equally spaced grid for $\lambda$ is defined by the arguments \code{lower}, \code{upper} and \code{n} with obvious meaning. Default values, for these parameters can be seen by
\begin{Schunk}
\begin{Sinput}
R> loggammarob.control()
\end{Sinput}
\end{Schunk}
Once a Q$\tau$ estimate $\tilde{\boldsymbol{\theta}}$ is obtained, the WQ$\tau$ is easily computed by first evaluating a fixed set of scales
\begin{equation*}
\tilde{\sigma}_j^2 = v^2(\tilde{\boldsymbol{\theta}}, u_{n,j})
\end{equation*}  
then the IRWLS is used with $\tilde{\boldsymbol{\theta}}$ as starting value and $r_j/\tilde{\sigma}_j$ in place of $r_j$.

\subsection{Computation of FIWL and 1SWL}

Weights need to be evaluated for both 1SWL and FIWL. To compute the kernel density estimate $f^\ast(y)$, \pkg{robustloggamma} uses the function \code{density} with \code{kernel="gaussian"}, \code{cut=3}, and \code{n=512}. The smoothed model $f_{\boldsymbol{\theta}}^\ast(y)$ is approximated by 
\begin{equation*}
\frac{1}{K} \sum_{k=1}^K k(y,y_k,h),
\end{equation*}
where $y_k$ is the quantile of order $(k-0.5)/K$ of $F_{\tilde{\boldsymbol{\theta}}}$. 
The bandwidth $h$ is adaptively fixed to \code{bw} times the actual value of $\sigma$ and $K$ is controlled by the argument \code{subdivisions}. The RAF is fixed by \code{raf} among several choices: 
\code{"NED"} (negative exponential disparity), 
\code{"GKL"} (generalized Kullback-Leibler), 
\code{"PWD"} (power divergence measure), 
\code{"HD"}  (Hellinger distance), 
\code{"SCHI2"} (symmetric Chi-Squared distance),
and \code{tau} selects the particular member of the family in case of \code{"GKL"} and \code{"PWD"}. Finally, weights smaller than \code{minw} are set to zero.

For 1SWL, $\mathbf{J} = \int w(y, \tilde{\boldsymbol{\theta}}) \nabla \mathbf{z}(y, \tilde{\boldsymbol{\theta}}) \ dF_{\tilde{\boldsymbol{\theta}}}(y)$ is approximated by
\begin{equation*}
\frac{1}{K} \sum_{k=1}^K w(y_k, \tilde{\boldsymbol{\theta}}) \nabla \mathbf{z}(y_k, \tilde{\boldsymbol{\theta}}) \ .
\end{equation*}
Here $K$ is controlled by \code{nexp}. Furthermore, the step can be multiplied by the \code{step} argument (with default $1$).


\section{An illustration}

We illustrate the use of  \pkg{robustloggamma} with the help of the data set \code{drg2000} included in the package. The data refer to $70323$ stays that were observed in year $2000$ in a group of Swiss hospitals within a pilot study aimed at the implementation of a diagnosis-related grouping (DRG) system. DRG systems are used in modern hospital management to classify each individual stay into a group according to the patient characteristics. The classification rules are defined so that the groups are as homogeneous as possible with respect to clinical criteria (diagnoses and procedures) and to resource consumption. A mean cost of each group is usually estimated yearly with the help of available data about the observed stays on a national basis. This cost is then assigned to each stay in the same group and used for reimbursement and budgeting.

Cost distributions are typically skewed and often contain outliers. When a small number of outliers are observed, the classical estimates of the mean can be much different than when none is observed. And since the values and the frequency of outliers fluctuate from year to year, the classical mean cost is unreliable. Not surprisingly, since many DRGs must routinely be inspected each year, automatic outlier detection is a recurrent hot topic for discussion among hospital managers.

The data set has four variables:
\code{LOS} length of stay, 
\code{Cost} cost of stay in Swiss francs, 
\code{APDRG} DRG code (according to the ``All Patients DRG'' system) 
and \code{MDC} Major diagnostic category. Packages \code{xtable} \citep{xtable} and \code{lattice} \citep{lattice} will be used during the illustration.
\begin{Schunk}
\begin{Sinput}
R> require("xtable")
R> require("lattice")
R> data("drg2000")
\end{Sinput}
\end{Schunk}
We will analyse the variable \code{Cost} on the logarithmic scale for the following four DRGs:
\begin{center}
\begin{tabular}{lr}
AP-DRG & Description \\
185 & Dental \& oral dis exc exctract \& restorations, Age > 17 \\
222 & Knee procedures w/o cc \\
237 & Sprain, strain, disloc of hip, pelvis, thigh \\
360 & Vagina, Cervix \& Vulva Procedures \\
\end{tabular}  
\end{center}


\begin{Schunk}
\begin{Sinput}
R> APDRG <- c(185, 222, 237, 360)
R> index <- unlist(sapply(APDRG, function(x) which(drg2000$APDRG==x)))
R> DRG <- drg2000[index,]
\end{Sinput}
\end{Schunk}

\begin{figure}
  \centerline{
     \includegraphics[width=0.7\textwidth]{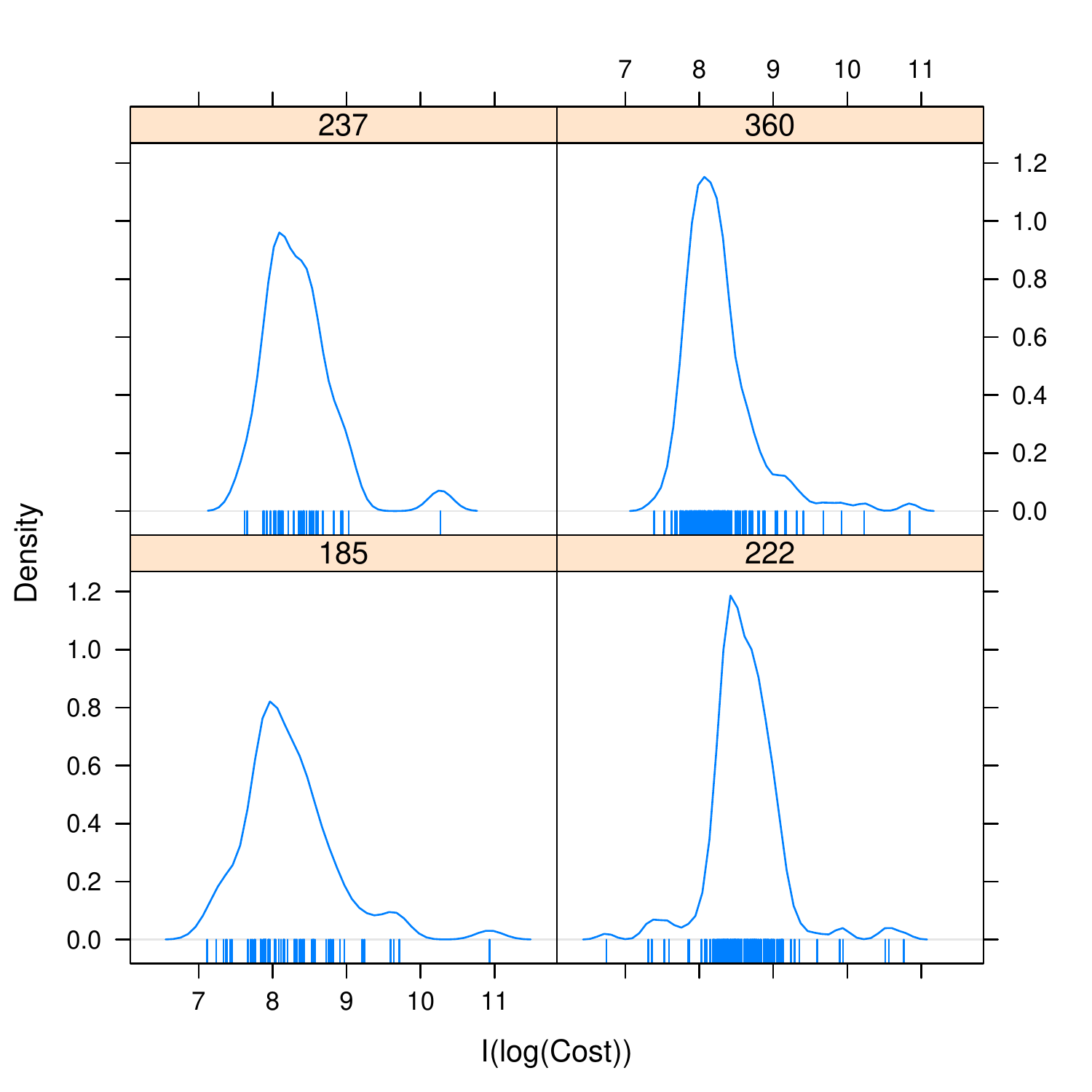}
  }
  \caption{Estimated densities of \code{log(Cost)} for selected DRGs.}
  \label{fig:data}
\end{figure}

Figure~\ref{fig:data}, obtained with the following code, shows a density plot for each selected DRG:

\begin{Schunk}
\begin{Sinput}
R> print(densityplot(~I(log(Cost)) | factor(APDRG), data=DRG, 
+                   plot.points="rug", ref=TRUE))
\end{Sinput}
\end{Schunk}

Summary statistics are:
\begin{Schunk}
\begin{Sinput}
R> lapply(split(DRG$Cost,DRG$APDRG), summary)
\end{Sinput}
\begin{Soutput}
$`185`
   Min. 1st Qu.  Median    Mean 3rd Qu.    Max. 
   1228    2645    3462    5059    5047   55770 

$`222`
   Min. 1st Qu.  Median    Mean 3rd Qu.    Max. 
  849.2  4362.0  5288.0  6307.0  6801.0 47240.0 

$`237`
   Min. 1st Qu.  Median    Mean 3rd Qu.    Max. 
   2038    3144    4100    4987    5169   28780 

$`360`
   Min. 1st Qu.  Median    Mean 3rd Qu.    Max. 
   1620    2863    3502    4680    4428   51160 
\end{Soutput}
\end{Schunk}
A comparison between classical and robust measures of spread, indicates important differences:
\begin{Schunk}
\begin{Sinput}
R> lapply(split(DRG$Cost,DRG$APDRG), function(x) c(sd(x), mad(x)))
\end{Sinput}
\begin{Soutput}
$`185`
[1] 6894.890 1374.874

$`222`
[1] 5095.599 1648.444

$`237`
[1] 4490.546 1494.891

$`360`
[1] 5066.084 1075.537
\end{Soutput}
\end{Schunk}
The differences are due to the presence of outliers. Therefore, it is convenient to analyze the data with the help of robust methods. 
To begin with, we use the function \code{loggammarob} to fit a generalized loggamma model to sample \code{APDRG=185}. This function provides robust estimates of the parameters $\mu$ (location), $\sigma$ (scale), and $\lambda$ (shape) using the default method 1SWL:
\begin{Schunk}
\begin{Sinput}
R> Cost185 <- sort(DRG$Cost[DRG$APDRG==185])
R> est185 <- loggammarob(log(Cost185))
R> est185
\end{Sinput}
\begin{Soutput}
Call:
loggammarob(x = log(Cost185))

Location:  8.04  Scale:  0.4944  Shape:  -0.6437  E(exp(X)):  4381
\end{Soutput}
\end{Schunk}
In addition, a \code{summary} method is available to calculate confidence intervals (based on the Wald statistics) for the parameters and for selected model quantiles (argument \code{p}).
\begin{Schunk}
\begin{Sinput}
R> summary(est185, p=c(0.9, 0.95, 0.99))
\end{Sinput}
\begin{Soutput}
Call:
summary.loggammarob(object = est185, p = c(0.9, 0.95, 0.99))

Location:  8.04 s.e.  0.09841 
( 7.847 ,  8.233 ) 
95 percent confidence interval

Scale:  0.4944 s.e.  0.05071 
( 0.395 ,  0.5938 ) 
95 percent confidence interval

Shape:  -0.6437 s.e.  0.3005 
( -1.233 ,  -0.05467 ) 
95 percent confidence interval

Mean(exp(X)):  4381 s.e.  426.7 
( 3545 ,  5218 ) 
95 percent confidence interval


Quantile of order  0.9 :  8.932 s.e.  0.2337 
( 8.474 ,  9.39 ) 
95 percent confidence interval


Quantile of order  0.95 :  9.2 s.e.  0.343 
( 8.528 ,  9.873 ) 
95 percent confidence interval


Quantile of order  0.99 :  9.774 s.e.  0.6505 
( 8.499 ,  11.05 ) 
95 percent confidence interval




Robustness weights: 
 54 weights are ~= 1. The remaining 15 ones are summarized as
   Min. 1st Qu.  Median    Mean 3rd Qu.    Max. 
0.05591 0.74150 0.88590 0.77530 0.99800 0.99890 
\end{Soutput}
\end{Schunk}
The function also provides the robust weights that allow outlier's identification. For instance:
\begin{Schunk}
\begin{Sinput}
R> which(est185$weights < 0.1)
\end{Sinput}
\begin{Soutput}
[1] 1
\end{Soutput}
\end{Schunk}
reports the indices of the observations with weights smaller than $0.1$, which in this case is the first one only.

Robust tests on one or more parameters can be performed by means of the weighted Wald test described in \citet{agostinellimarkatou2001}. For this purpose, we use the function \code{loggammarob.test}. For instance, we test the hypothesis that the shape parameter is equal to zero, i.e., that the lognormal model is an acceptable one:
\begin{Schunk}
\begin{Sinput}
R> loggammarob.test(est185, lambda=0)
\end{Sinput}
\begin{Soutput}
	Weighted Wald Test based on oneWL

data:  
ww = 4.5876, df = 1, p-value = 0.0322
alternative hypothesis: true shape is not equal to 0
95 percent confidence interval:
 -1.23270096 -0.05466982
sample estimates:
[1] -0.6436854
\end{Soutput}
\end{Schunk}
To test the hypothesis that the location is zero and the scale is one we use:
\begin{Schunk}
\begin{Sinput}
R> loggammarob.test(est185, mu=0, sigma=1)
\end{Sinput}
\end{Schunk}
however, in these situations, the confidence intervals are not calculated.

The default estimation method in \code{loggammarob} is 1SWL (one-step weighted likelihood). However, alternative estimates are made available: Q$\tau$, WQ$\tau$, WL, and ML. Q$\tau$ and WQ$\tau$, typically used as starting values for the weighted likelihood procedures, are obtained as follows:
\begin{Schunk}
\begin{Sinput}
R> qtau185 <- summary(loggammarob(log(Cost185), method="QTau"))
R> wqtau185 <- summary(loggammarob(log(Cost185), method="WQTau"))
\end{Sinput}
\end{Schunk}
The fully iterated weighted likelihood (FIWL) and the one-step weighted likelihood estimates (1SWL) are obtained as follows:
\begin{Schunk}
\begin{Sinput}
R> fiwl185 <- summary(loggammarob(log(Cost185), method="WL"))
R> oswl185 <- summary(loggammarob(log(Cost185), method="oneWL"))
\end{Sinput}
\end{Schunk}
The maximum likelihood estimate is also available:
\begin{Schunk}
\begin{Sinput}
R> ml <- summary(loggammarob(log(Cost185), method="ML"))
\end{Sinput}
\end{Schunk}
We now compare the four samples. For this purpose, the function \code{analysis} available in the Supplemental Material must be loaded before running the next command.

\begin{Schunk}
\begin{Sinput}
R> results <- sapply(APDRG, function(x) analysis(APDRG=x, data=DRG), 
+              simplify=FALSE)
\end{Sinput}
\end{Schunk}

\begin{figure}
  \centerline{
     \includegraphics[width=0.7\textwidth]{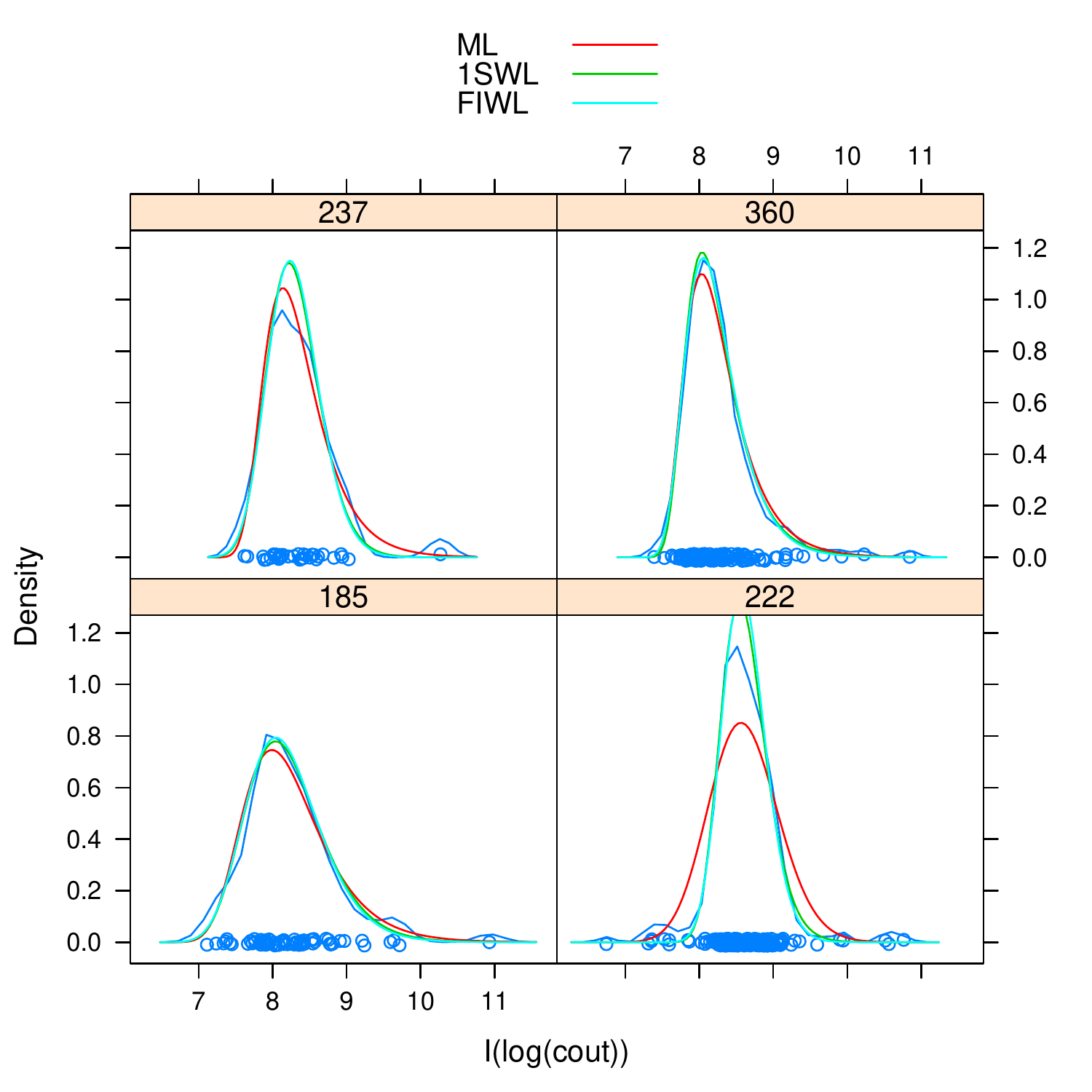}
  }
  \caption{Estimated models provided by ML, 1SWL and FIWL for the four data sets.}
  \label{fig:plots}
\end{figure}
\begin{figure}
  \centerline{
     \includegraphics[width=0.7\textwidth]{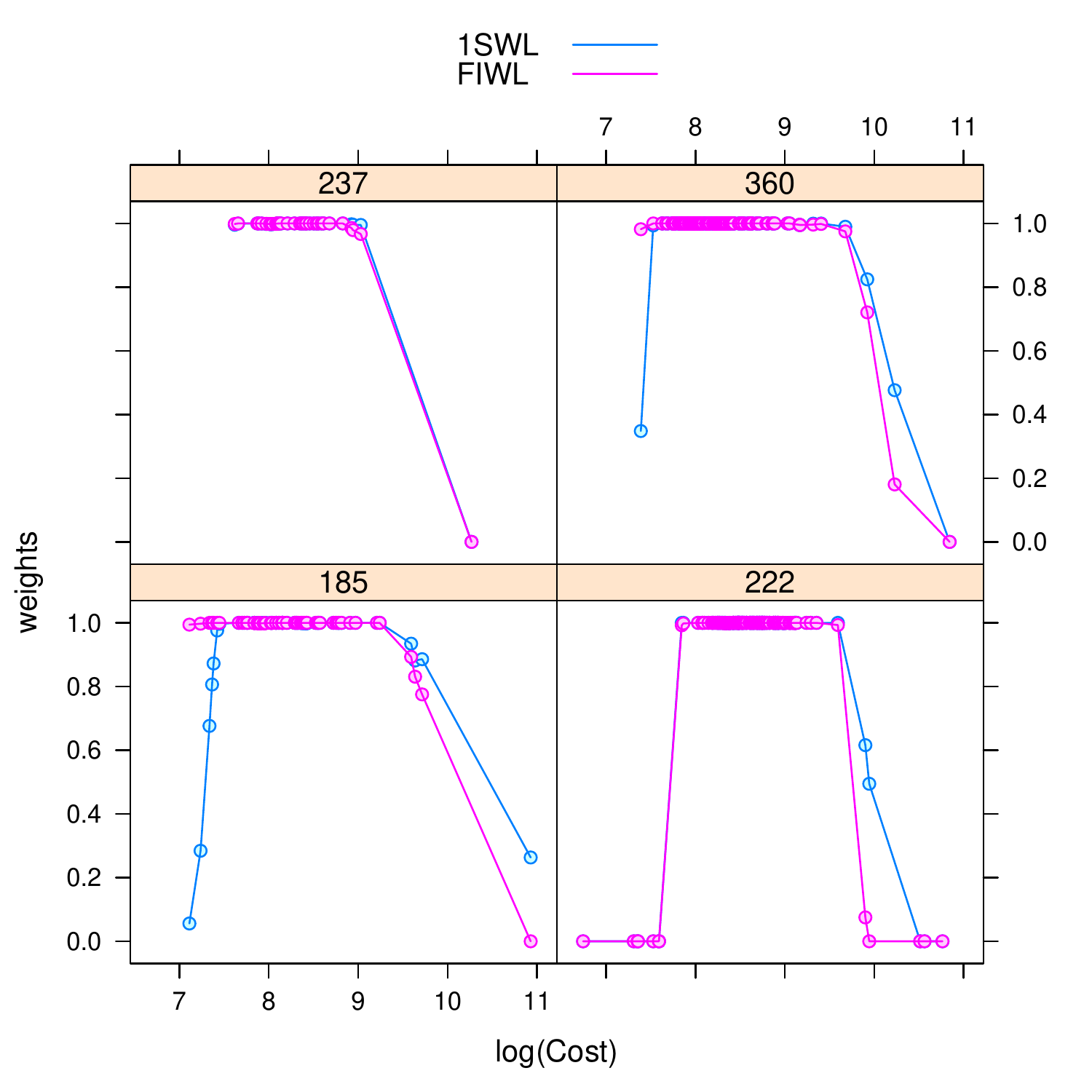}
  }
  \caption{Weights provided by 1SWL and FIWL.}
  \label{fig:weights}
\end{figure}

Three estimated densities provided by ML, FIWL, and 1SWL are shown in Figure~\ref{fig:plots}. The robust parameter estimates, their estimated standand errors, and their confidence intervals are shown in Table~\ref{tab:est}. The table was obtained with the help of the the function \code{maketable} available in the Supplemental Material. 
\begin{table}
\begin{center}
  \resizebox{\textwidth}{!} {
  \begin{tabular}{ll|rrrr|rrrr|rrrr|}
 DRG & & \multicolumn{4}{c|}{ML} & \multicolumn{4}{c|}{1SWL} & \multicolumn{4}{c|}{FIWL} \\
  \hline  
  & & $\mu$ & $\sigma$ & $\lambda$ & $\eta$ & $\mu$ & $\sigma$ & $\lambda$ & $\eta$ & $\mu$ & $\sigma$ & $\lambda$ & $\eta$ \\  
  \hline  
 185 & est & 7.989 & 0.501 & -0.892 & 4837 & 8.04 & 0.494 & -0.644 & 4381 & 8.043 & 0.489 & -0.574 & 4232 \\ 
    & se & 0.101 & 0.056 & 0.305 & 657 & 0.098 & 0.051 & 0.301 & 427 & 0.095 & 0.048 & 0.295 & 372 \\ 
    & left & 7.792 & 0.392 & -1.49 & 3550 & 7.847 & 0.395 & -1.233 & 3545 & 7.856 & 0.394 & -1.153 & 3504 \\ 
    & right & 8.186 & 0.61 & -0.295 & 6124 & 8.233 & 0.594 & -0.055 & 5218 & 8.23 & 0.583 & 0.005 & 4961 \\ 
   \hline
222 & est & 8.563 & 0.467 & -0.197 & 6152 & 8.544 & 0.295 & -0.493 & 5836 & 8.566 & 0.289 & -0.308 & 5747 \\ 
    & se & 0.054 & 0.025 & 0.179 & 237 & 0.035 & 0.017 & 0.183 & 156 & 0.035 & 0.016 & 0.184 & 138 \\ 
    & left & 8.457 & 0.419 & -0.549 & 5687 & 8.475 & 0.261 & -0.852 & 5531 & 8.498 & 0.257 & -0.669 & 5475 \\ 
    & right & 8.67 & 0.515 & 0.155 & 6618 & 8.613 & 0.329 & -0.135 & 6141 & 8.634 & 0.321 & 0.054 & 6018 \\ 
   \hline
237 & est & 8.139 & 0.349 & -1.05 & 4836 & 8.221 & 0.344 & -0.432 & 4300 & 8.236 & 0.343 & -0.329 & 4266 \\ 
    & se & 0.103 & 0.059 & 0.455 & 596 & 0.095 & 0.046 & 0.423 & 308 & 0.095 & 0.045 & 0.425 & 290 \\ 
    & left & 7.936 & 0.234 & -1.942 & 3667 & 8.034 & 0.254 & -1.261 & 3697 & 8.05 & 0.256 & -1.162 & 3697 \\ 
    & right & 8.341 & 0.465 & -0.158 & 6005 & 8.408 & 0.434 & 0.396 & 4902 & 8.422 & 0.431 & 0.504 & 4835 \\ 
   \hline
360 & est & 8.034 & 0.324 & -1.191 & 4444 & 8.038 & 0.304 & -1.137 & 4235 & 8.055 & 0.318 & -0.974 & 4142 \\ 
    & se & 0.049 & 0.029 & 0.237 & 280 & 0.046 & 0.026 & 0.234 & 228 & 0.046 & 0.026 & 0.221 & 201 \\ 
    & left & 7.938 & 0.268 & -1.655 & 3896 & 7.949 & 0.252 & -1.595 & 3789 & 7.965 & 0.267 & -1.407 & 3748 \\ 
    & right & 8.129 & 0.38 & -0.727 & 4993 & 8.127 & 0.356 & -0.679 & 4681 & 8.145 & 0.368 & -0.542 & 4535 \\   \hline
  \end{tabular}
  }
  \caption{Estimates, standard errors and confidence intervals for the four samples.}
  \label{tab:est}
\end{center}
\end{table}
We note that outliers are present in all samples (Figure~\ref{fig:weights}). However, for \code{APDRG=185} and \code{APDRG=360} they do not have a significant impact on the estimates and the associated inferences. On the contrary, for \code{APDRG=222} the outliers markedly inflate the ML scale estimate and, for \code{APDRG=237} a single outlier has a great impact on the ML estimate of lambda. For this sample, the robust parameter estimates and their confidence intervals suggest that the lognormal density is a possible model. To visualize the weights, we use the following code:
\begin{Schunk}
\begin{Sinput}
R> weights <- function(x, results) {
+   os <- results[[x]]$os
+   wl <- results[[x]]$wl
+   ans <- t(cbind(os$weights, wl$weights, wl$data, x))
+   return(ans)
+ }
R> w <- as.data.frame(matrix(unlist(sapply(1:4, 
+   function(x) weights(x, results=results))), ncol=4, byrow=TRUE))
R> colnames(w) <- c("OSWL", "FIWL", "data", "drg")
R> w$drg <- factor(w$drg, labels=APDRG)
R> lattice.theme <- trellis.par.get()
R> col <- lattice.theme$superpose.symbol$col[1:2]
R> print(xyplot(OSWL+FIWL~data | drg, data=w, type="b",
+   col=col, pch=21, key = list(text=list(c("1SWL", "FIWL")), 
+   lines=list(col=col)), xlab="log(Cost)", ylab="weights"))
\end{Sinput}
\end{Schunk}

Is the two-parameter gamma distribution an acceptable model for the four samples? To answer this question, we test the hypothesis $\sigma=\lambda$. We use the function \code{loggammarob.wilks} with weights provided by ML, 1SWL and, FIWL. The results are summarized in Table~\ref{tab:wilks} provided by the function \code{extractwilks} reported in the Supplemental Material. The hypothesis is always strongly rejected for DRGs 185, 222, 360. For DRG 237, the presence of outliers leads ML to stronlgy reject the hypothesis, while the robust methods accept it ($\mu_0$ and $\sigma_0=\lambda_0$ are the estimated parameters under the null hypothesis).

\begin{Schunk}
\begin{Sinput}
R> wilks <- extractwilks(results)
R> wilks <- cbind(c("185", rep(" ", 3), "222", rep(" ", 3), 
+                    "237", rep(" ", 3), "360", rep(" ", 3)),
+                rep(c("Statistic", "p-value", "$\\mu_0$", 
+                "$\\sigma_0$"),4), wilks)
R> xwilks <- xtable(wilks)
\end{Sinput}
\end{Schunk}

\begin{figure}
  \centerline{
     \includegraphics[width=0.7\textwidth]{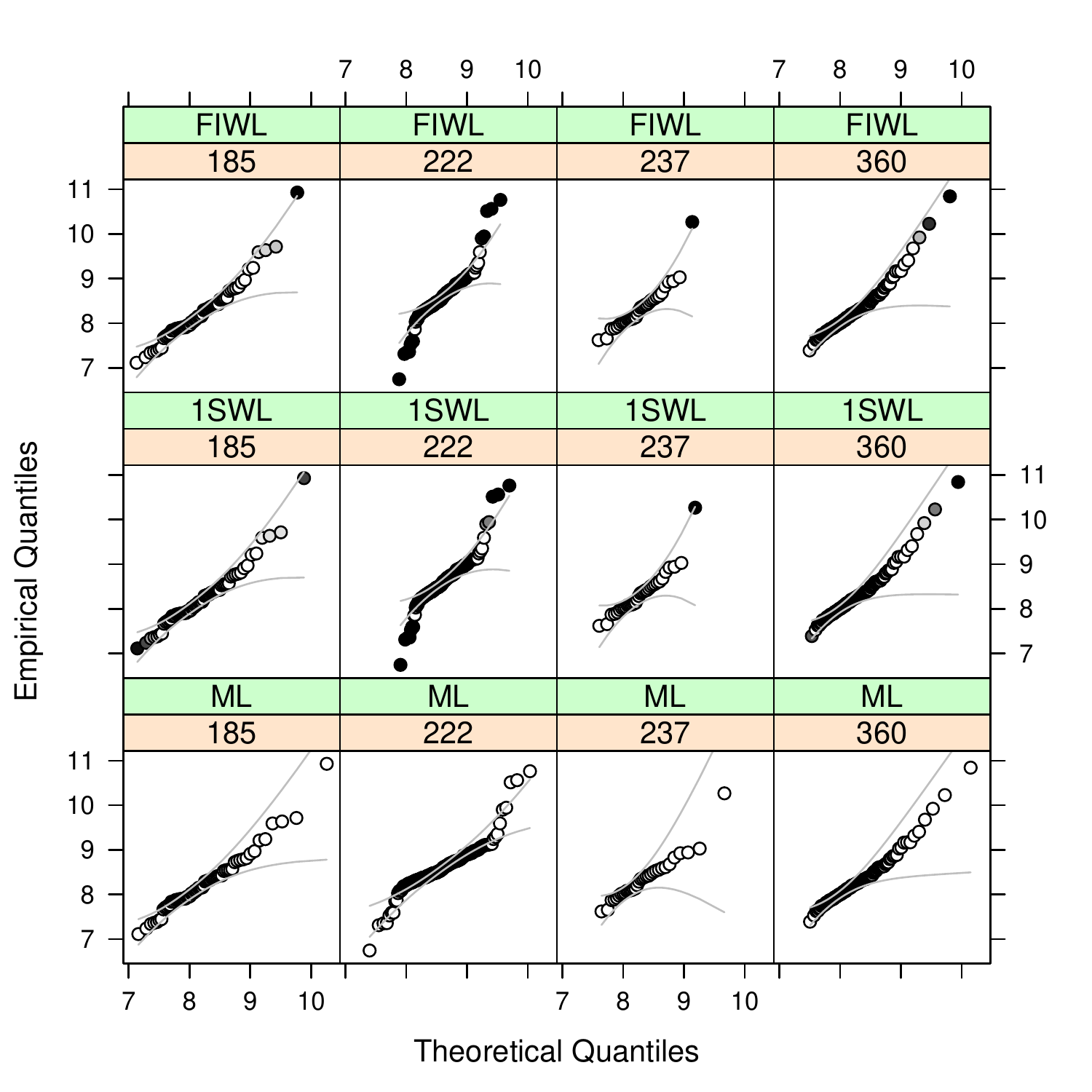}
  }
  \caption{Q-Q plots and $90\%$ confidence bands (grey line). Darker points are associated with smaller weights.}
  \label{fig:qqplot}
\end{figure}

\begin{table}
\begin{center}
  \begin{tabular}{ll|rrr|}
  \hline  
  DRG & & ML & 1SWL & FIWL \\
  \hline
 185 & Statistic & 45.882 & 29.1741 & 15.3918 \\ 
    & p-value & 0 & 0 & 1e-04 \\ 
    & $\mu_0$ & 8.5288 & 8.4277 & 8.3497 \\ 
    & $\sigma_0$ & 0.7281 & 0.5949 & 0.5502 \\ 
   \hline
222 & Statistic & 52.3126 & 24.7645 & 10.8147 \\ 
    & p-value & 0 & 0 & 0.001 \\ 
    & $\mu_0$ & 8.7494 & 8.6712 & 8.6563 \\ 
    & $\sigma_0$ & 0.5172 & 0.3208 & 0.2984 \\ 
   \hline
237 & Statistic & 23.1627 & 1.7348 & 1.7679 \\ 
    & p-value & 0 & 0.1878 & 0.1836 \\ 
    & $\mu_0$ & 8.5146 & 8.3583 & 8.3567 \\ 
    & $\sigma_0$ & 0.5548 & 0.3527 & 0.3518 \\ 
   \hline
360 & Statistic & 123.933 & 75.6255 & 62.5466 \\ 
    & p-value & 0 & 0 & 0 \\ 
    & $\mu_0$ & 8.4512 & 8.3543 & 8.3363 \\ 
    & $\sigma_0$ & 0.591 & 0.4654 & 0.4481 \\   \hline
  \end{tabular}
  \caption{Weighted Wilks test for the hypothesis $\sigma=\lambda$.}
  \label{tab:wilks}
\end{center}
\end{table}
Finally, we draw Q-Q plots based on ML, 1SWL and FIWL (Figure~\ref{fig:qqplot}) for the four data sets. Darker points are associated with smaller weights. $90\%$ confidence bands are provided to check the adequacy of the model to the data.
\begin{Schunk}
\begin{Sinput}
R> quant <- function(x, method, results) {
+   res <- results[[x]][[method]]
+   n <- length(res$data)
+   q <- qloggamma(p=ppoints(n), mu=res$mu, sigma=res$sigma, lambda=res$lambda)  
+   qconf <- summary(res, p=ppoints(n), conf.level=0.90)$qconf.int
+   ans <- t(cbind(q, qconf, res$data, res$weights, x, method))
+   return(ans)
+ }
R> q1 <- matrix(unlist(sapply(1:4, 
+   function(x) quant(x, method=1, results=results))), 
+   ncol=7, byrow=TRUE)
R> q2 <- matrix(unlist(sapply(1:4, 
+   function(x) quant(x, method=2, results=results))), 
+   ncol=7, byrow=TRUE)
R> q3 <- matrix(unlist(sapply(1:4, 
+   function(x) quant(x, method=3, results=results))), 
+   ncol=7, byrow=TRUE)
R> q <- as.data.frame(rbind(q1,q2,q3))
R> colnames(q) <- c("q", "qlower", "qupper", "Cost", 
+   "weights", "drg", "method")
R> q$drg <- factor(q$drg, labels=APDRG)
R> q$method <- factor(q$method, labels=c("ML", "1SWL", "FIWL"))
\end{Sinput}
\end{Schunk}

\begin{Schunk}
\begin{Sinput}
R> print(xyplot(Cost~q | drg+method, data=q, xlab="Theoretical Quantiles",
+   ylab="Empirical Quantiles", fill.color=grey(q$weights), q=q, 
+   panel=function(x, y, fill.color, ..., subscripts, q) {
+     fill=fill.color[subscripts]
+     q=q[subscripts,]
+     panel.xyplot(x, y, pch=21, fill=fill, col="black", ...)
+     panel.xyplot(x, y=q$qupper, type="l", col="grey75")
+     panel.xyplot(x, y=q$qlower, type="l", col="grey75")  
+   } 
+ ))
\end{Sinput}
\end{Schunk}

\section{Acknowledgments}
All statistical analysis were performed on SCSCF (\href{www.dais.unive.it/scscf}{www.dais.unive.it/scscf}), a multiprocessor cluster system owned by Ca' Foscari University of Venice running under GNU/Linux.

This research was partially supported by the Italian-Argentinian project ``Metodi robusti per la previsione del costo e della durata della degenza ospedaliera'' funded in the collaboration program MINCYT-MAE AR14MO6.

V\'{\i}ctor Yohai research was partially supported by Grants 20020130100279 from Universidad of Buenos Aires, PIP 112-2008-01-00216 and 112-2011-01-00339 from CONICET and PICT 2011-0397 from ANPCYT.


\appendix

\section[Tau scale regression]{$\tau$ scale regression}
\label{app:tauscaleregression}
In this appendix we briefly review the definition of $\tau$ scale and $\tau$ regression. For a detailed description see \citet{maronnamartinyohai2006}.

Let $\rho $ be a function satisfying the following properties: \textbf{A}: (i) $\rho(0)=0$; (ii) $\rho$ is even; (iii) if $|x_1| < |x_2|$, then $\rho(x_1) \leq \rho(x_2)$; (iv) $\rho$ is bounded; (v) $\rho$ is continuous.

Then, an \textit{M scale} \citep{huber1981} based on $\rho$ is defined by the value $s$ satisfying
\begin{equation*}
\frac{1}{n}\sum\limits_{j=1}^n\rho \left( \frac{u_j}{s} \right) = b \ ,
\end{equation*}
where $b$ is a given scalar and $0 < b < a =\sup \rho$. \citet{yohaizamar1988} introduce the family of $\tau$ scales. A $\tau$ \textit{scale} is based on two functions $\rho_1$ and $\rho_2$ satisfying properties \textbf{A} and such that $\rho_2 \leq \rho_1$. To define a $\tau$ scale, one considers an M scale $s_1^2(\mathbf{u})$ based on $\rho_1$ then, the $\tau$ scale is given by
\begin{equation*}
\tau^2(\mathbf{u}) = s_1^2(\mathbf{u}) \frac{1}{n} \sum_{j=1}^n \rho_2 \left( \frac{u_j} {s_1(\mathbf{u})} \right) \ .
\end{equation*}
$\tau$ scale estimators can be extended easily to the linear regression case. Let us consider the regression model
\begin{equation*}
y_j = \boldsymbol{\beta}^t \mathbf{x}_j + e_j \ , \ 1 \leq j \leq n \ ,
\end{equation*}
where $\boldsymbol{\beta} = (\beta_1, \cdots, \beta_p)^t$ and $\mathbf{x}_j = (x_{j1}, \cdots, x_{jp})^t$. For a given $\boldsymbol{\beta}$, let $r_j(\boldsymbol{\beta}) = y_j - \boldsymbol{\beta}^t \mathbf{x}_j$ be the corresponding residuals. The scale $\tau^2(r_1(\boldsymbol{\beta}), \cdots, r_n(\boldsymbol{\beta}))$ may be considered as a measure of goodness of fit. Based on this remark, \citet{yohaizamar1988} define robust estimators of the coefficients of a regression model by
\begin{equation*}
\hat{\boldsymbol{\beta}} = \arg\min_{\boldsymbol{\beta}} \tau \left( r_1(\mathbf{%
\beta }), \cdots, r_n(\mathbf{\beta })\right) \ .
\end{equation*}
These estimators are called $\tau$ \textit{regression estimators}. If $a/b=0.5$, the $\tau$ estimators have breakdown point (bdp) close to $50\%$ \citep{yohaizamar1988}. Moreover, we note that, if $\rho_2(u)=u^2$, $\tau^2(u_1, \cdots, u_n)=\text{ave}(u_j^2)$ and then the regression $\tau$ estimator coincides with the least squares estimator. Therefore, taking as $\rho_2$ a bounded function close to the quadratic function, the regression $\tau$ estimators can be made arbitrarily efficient for normal errors. If the errors $e_j$ are heteroscedastic with variances proportional to $\sigma_j^2$, the efficiency of $\hat{\boldsymbol{\beta}}$
can be improved by means of a weighted procedure. A \textit{regression weighted} $\tau$ \textit{estimator} is given by 
\begin{equation*}
\hat{\boldsymbol{\beta}} = \arg\min_{\boldsymbol{\beta}} \tau \left( r_1^{\ast}(\boldsymbol{\beta}), \cdots, r_n^{\ast}( \boldsymbol{\beta}) \right) \ ,
\end{equation*}
where $r_j^{\ast}(\boldsymbol{\beta}) = r_j(\boldsymbol{\beta})/\sigma_j$.

Usually, one chooses $\rho_1$ and $\rho_2$ in the \textit{Tukey biweight family}
\begin{equation*}
\rho(u, c) = \left\{ 
\begin{array}{lll}
3(u/c)^2-3(u/c)^4+(u/c)^6 & \text{if} & |u| \leq c \ , \\ 
1 & \text{if} & |u| > c \ ,
\end{array}
\right.
\end{equation*}
using two values $c_1$ and $c_2$ of the tuning parameter $c$. For example, one can take $c_1=1.548$ and $c_2=6.08$. With $b=0.5$, these values yield regression estimators with breakdown point $0.5$ and normal efficiency of $95\%$.

\section{Residual adjustment functions}
\label{app:RAF}

The literature provides several proposals for selecting the RAF. In the following, we recall two of them. The RAF based on the \textit{power divergence measure} (PWD) \citep{cressie1984,cressie1988}, is given by
\begin{equation*}
A_{pdm}(\delta, \tau) = \left\{
\begin{array}{ll}
\tau \left( (\delta + 1)^{1/\tau} - 1 \right) & \tau < \infty \\
\log(\delta + 1) & \tau \rightarrow \infty 
\end{array}
\right.
\end{equation*}
Special cases are \textit{maximum likelihood} ($\tau = 1$), \textit{Hellinger distance} ($\tau = 2$), \textit{Kullback--Leibler divergence} ($\tau \rightarrow \infty$), and \textit{Neyman's Chi--Square} ($\tau=-1$). The RAF based on the \textit{generalized Kullback--Leibler divergence} (GKL; \citet{parkbasu2003}) is given by:
\begin{equation*}
A_{gkl}(\delta, \tau) = \frac{\log(\tau \delta + 1)}{\tau} \qquad 0 \le \tau \le 1 \ .
\end{equation*}
Special cases are maximum likelihood ($\tau \rightarrow 0$) and Kullback--Leibler divergence ($\tau = 1$). This RAF can be interpreted as a linear combination between the likelihood divergence and the Kullback--Leibler divergence. A further example is the RAF corresponding to the \textit{negative exponential dsparity} (NED; \citep{lindsay1994})
\begin{equation*}
A(\delta) = 2 - (2 + \delta) \exp(-\delta)
\end{equation*}
which, for discrete models is second order efficient. 

\section{Reference chart}
\label{app:help}

Hereafter we provide the reference chart for the main function \code{loggammarob}. The usage has the following form
\begin{Schunk}
\begin{Sinput}
R> loggammarob(x, start=NULL, weights = rep(1, length(x)),
+   method=c("oneWL", "WQTau", "WL", "QTau", "ML"), control, ...)
\end{Sinput}
\end{Schunk}
where

\noindent \code{x} is a numeric vector, which contains the data.

\noindent \code{start} is \code{NULL} or a numeric vector containing the starting values of location, scale, and shape to be used when method is "WL", "oneWL" and "ML".  Method "QTau" does not require starting values. When start is \code{NULL}, the methods "QTau" and "WQTau" are called in a series to compute the starting values.

\noindent \code{weights} is a numeric vector containing the weights for method "QTau".

\noindent \code{method} is a character string to select the method. The default is "oneWL" (one step weighted likelihood estimate starting from "WQTau"). 
Others available methods are 
"WL" (fully iterated weighted likelihood estimate),
"WQTau" (weighted Q$\tau$  estimate), 
"QTau" (Q$\tau$ estimate), and 
"ML" (maximum likelihood estimate).

\noindent \code{control} is a list, which contains an object from the function loggammarob.control.

\noindent \code{...} further arguments that can be directly passed to the function.

\noindent The function returns an object of class 'loggammarob'. This is a list with the following components:

\noindent \code{mu}: location parameter estimate.

\noindent \code{sigma}: scale parameter estimate.

\noindent \code{lambda}: shape parameter estimate.

\noindent \code{eta}: estimate of E(exp(x)).

\noindent \code{weights}: the final weights.

\noindent \code{iterations}: number of iterations.
\end{document}